\documentclass[%
 reprint,
 amsmath,amssymb,
 physrev,
]{revtex4-2}

\usepackage{graphicx}
\usepackage{dcolumn}
\usepackage{bm}
\usepackage{hyperref}

\begin{document}

\title{Failure to Reproduce the Results of ``A new transferable interatomic potential for molecular dynamics simulations of borosilicate glasses''}

\author{Fran\c{c}ois-Xavier Coudert}
 \email{fx.coudert@chimieparistech.psl.eu}
 \homepage{https://www.coudert.name/}
\affiliation{Chimie ParisTech, PSL Research University, CNRS, Institut de Recherche de Chimie Paris, 75005 Paris, France}%

\date{\today}

\begin{abstract}
We reproduced the simulations described in Wang et al [\emph{J. Non-Cryst. Sol.}, 498 (\textbf{2018}) 294--304] and found we could not obtain the results reported. The root cause was identified to be incorrect atom masses in the original simulation files. As a consequence, the potential does not reproduce the experimental glass density --- and presumably, other structural properties --- and should be considered with great caution.
\end{abstract}

\maketitle

Wang et al. reported in Ref.~[\onlinecite{Wang2018}] the development of a new empirical interatomic potential for borosilicate glasses. Based on a potential from Guillot and Sator,\cite{Guillot2007} that classical force field is composed of two-body Buckingham potential energy terms and Coulombic interactions:
\[ U_{ij}(r_{ij}) = \frac{z_i z_j}{r_{ij}} + A_{ij} \exp\left(-\frac{r_{ij}}{\rho_{ij}}\right) - \frac{C_{ij}}{r_{ij}^6} \]
The Wang potential used the parameters from Guillot and Sator, but since that did not contain any interaction parameters for B cations, the authors optimized the terms involving boron, i.e., the energy terms for B--O, B--B, and B--Si interactions. The parameters for these terms were determined by fitting on experimental data over the whole range of glass composition, from pure silicate to pure borate. The properties fitted include the coordination number of B cations (for the B--O parameters) and the evolution of density as a function of glass composition (for the B--B parameters). The authors then describe the structural properties of the glasses as obtained with their model.

\begin{figure}[b]
\includegraphics[width=\linewidth]{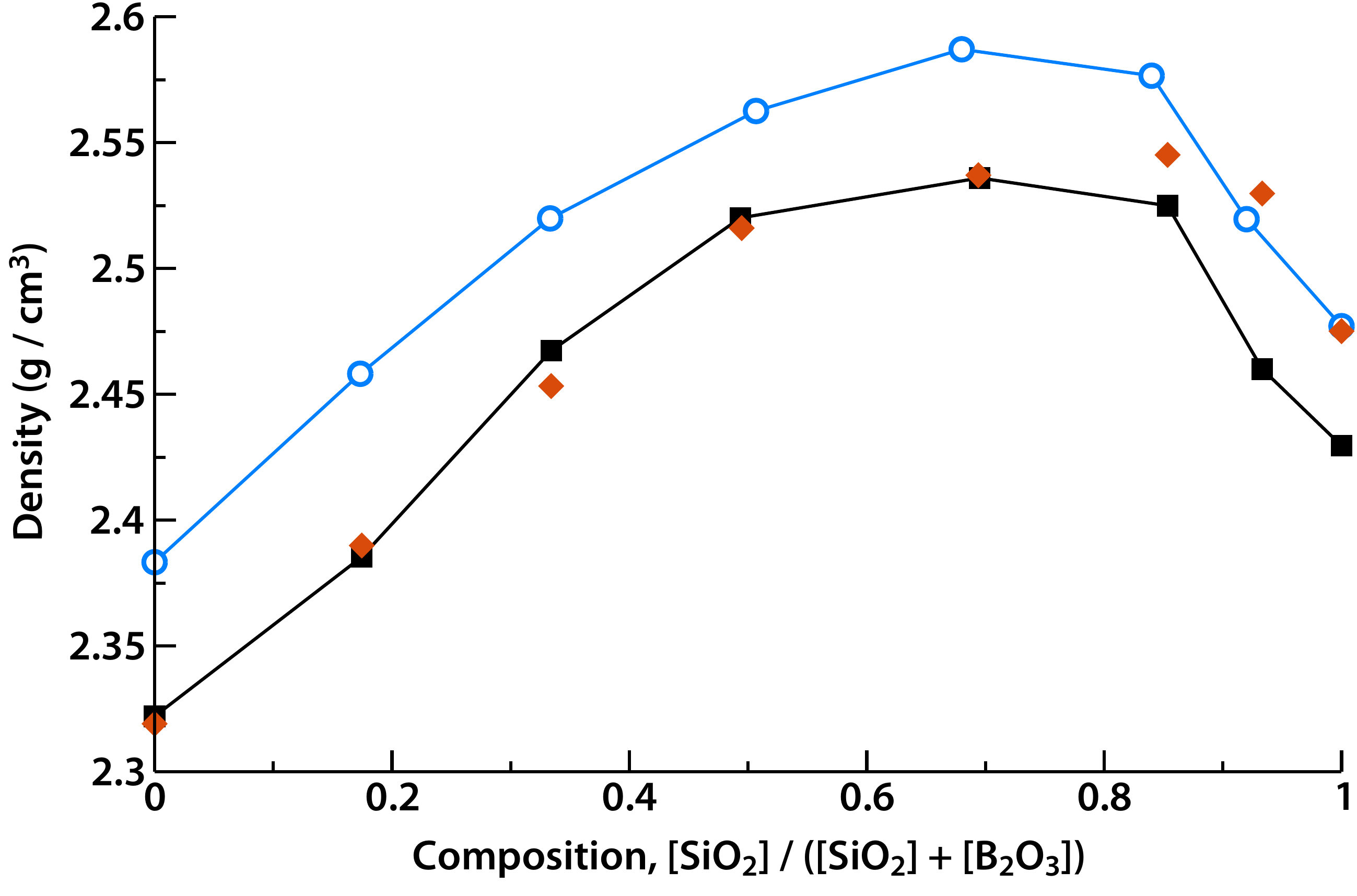}
\caption{Glass density as a function of composition, represented as fraction of SiO$_2$ (over the total amount of SiO$_2$ and B$_2$O$_3$) Red diamonds: experimental data.\cite{Smedskjaer2011} Black squares: results published in Ref.~[\onlinecite{Wang2018}]. Blue circles: our reproduction of the work. Lines are guides for the eye.
\label{fig:densities}}
\end{figure}

\section{Validity of the interatomic potential}

While trying to reproduce the work published in Ref.~[\onlinecite{Wang2018}], we could not reproduce the results in the authors' Figure 1, i.e., the evolution of density as a function of glass composition. We present the points from the original paper, and our reproduction, in Figure~\ref{fig:densities}. While the overall trend is somewhat similar, there is a systematic difference and we found densities that were systematically higher than reported, with a difference of up to 0.07~g/cm$^3$. The quantitative agreement with experimental data, reported by the authors, is not reproduced at all.

While an example LAMMPS\cite{Thompson2022} input file was provided in Supplementary information, the original configuration (LAMMPS data file) was not published. The author kindly shared starting configurations from the different systems, and we discovered that the original files had errors in the atomic masses of boron and silicon atoms. This impacted all simulations and results reported in Wang et al.

Moreover, its appears that the ``new transferable interatomic potential'' was itself optimized with the incorrect atomic masses. The potential parameters for the B--B terms were optimized to reproduce the experimental densities, but they took into account the wrong masses. Therefore, the potential should be reoptimized in the future before it is further used, especially the B--B parameters.

\section{Works impacted}

The paper by Wang et al. is higly cited, with 117 citations on Google Scholar to date. Having reviewing the citations, we think the following works are relying on the invalid potential for all or part of their molecular simulations:

\begin{itemize}

\item Wang, M.; Smedskjaer, M. M.; Mauro, J. C.; Bauchy, M. \href{https://doi.org/10.1063/1.5051746}{``Modifier Clustering and Avoidance Principle in Borosilicate Glasses: A Molecular Dynamics Study''}. \emph{J. Chem. Phys.}, \textbf{2019}, 150, 044502. The densities reported are the same as Ref.~[\onlinecite{Wang2018}], and it is safe to assume it uses input files with invalid atomic masses.

\item Kumar, R.; Jan, A.; Bauchy, M.; Krishnan, N. M. A. \href{https://doi.org/10.1111/jace.18013}{``Effect of Irradiation on the Atomic Structure of Borosilicate Glasses''}. \emph{J. Am. Ceramic Soc.}, \textbf{2021}, 104, 6194--6206. This paper has common co-authors with Ref.~[\onlinecite{Wang2018}], and because it does reproduce the same densities as Wang et al., it seems that it used incorrect atomic masses as well. All results should be taken with great caution.

\item Tuheen, M. I.; Deng, L.; Du, J. \href{https://doi.org/10.1016/j.jnoncrysol.2020.120413}{``A Comparative Study of the Effectiveness of Empirical Potentials for Molecular Dynamics Simulations of Borosilicate Glasses''}. \emph{J. Non-Cryst. Sol.}, \textbf{2021}, 553, 120413. The authors' conclusion is that the potential ``fails to provide the correct B coordination change for a wide range of compositions'', which may be linked to the optimization issues.

\item Fortino, M.; Berselli, A.; Stone-Weiss, N.; Deng, L.; Goel, A.; Du, J.; Pedone, A. \href{https://doi.org/10.1111/jace.16655}{``Assessment of Interatomic Parameters for the Reproduction of Borosilicate Glass Structures via DFT-GIPAW Calculations''}. \emph{J. Am. Ceramic Soc.}, \textbf{2019}, 102, 7225--7243. In this paper comparing different interatomic potentials, the authors used $(N, V, T)$ simulations at fixed experimental densities.

\item Bisbrouck, N.; Micoulaut, M.; Delaye, J. M.; Gin, S.; Angeli, F. \href{https://doi.org/10.1038/s41529-022-00268-8}{``Structure--Property Relationship and Chemical Durability of Magnesium-Containing Borosilicate Glasses with Insight from Topological Constraints''}. \emph{npj Mater. Degrad.}, \textbf{2022}, 6, 58. These simulations were performed at fixed density, in the $(N, V, T)$ and $(N, V, E)$ ensembles.

\item Zhai, C.; Zhong, Y.; Liu, J.; Zhang, J.; Zhu, Y.; Wang, M.; Yeo, J. \href{https://doi.org/10.1016/j.jnoncrysol.2021.121273}{``Customizing the Properties of Borosilicate Foam Glasses via Additions under Low Sintering Temperatures with Insights from Molecular Dynamics Simulations''}. \emph{J. Non-Cryst. Sol.}, \textbf{2022}, 576, 121273. This work includes a detailed discussion of the evolution of density and other physical properties as a function of boron content of the glasses, and is probably significantly impacted.

\item Lee, K.; Yang, Y.; Ding, L.; Ziebarth, B.; Mauro, J. C. \href{https://doi.org/10.1111/jace.18207}{``Effect of Pressurization on the Fracture Toughness of Borosilicate Glasses''}. \emph{J. Am. Ceramic Soc.}, \textbf{2021}, 105, 2536--2545. The Wang force field was compared to other potentials, and is found to lead to higher densities and different trends in Young's modulus. This appears to be linked to the parameterization issue.

\item Ding, L.; Lee, K.; Zhao, T.; Yang, Y.; Bockowski, M.; Ziebarth, B.; Wang, Q.; Ren, J.; Smedskjaer, M. M.; Mauro, J. C. \href{https://doi.org/10.1111/jace.17377}{``Atomic Structure of Hot Compressed Borosilicate Glasses''}. \emph{J. Am. Ceramic Soc.}, \textbf{2020}, 103, 6215--6225. For one of the systems (Boro33), this study features $(N, P, T)$ simulations. We note that the authors report that ``A computed pressure [\ldots] of $-1.8$~GPa is found in Boro33 glass (Bauchy's potential). The pressure at room temperature density is due to the choice of potential''. Moreover, because several authors are also co-authors of Ref.~[\onlinecite{Wang2018}], it is unclear whether the correct atomic masses were used in the simulations.

\item Lee, K.-H.; Yang, Y.; Ziebarth, B.; Mannstadt, W.; Davis, M. J.; Mauro, J. C. \href{https://doi.org/10.1016/j.jnoncrysol.2019.119736}{``Evaluation of Classical Interatomic Potentials for Molecular Dynamics Simulations of Borosilicate Glasses''}. \emph{J. Non-Cryst. Sol.}, \textbf{2020}, 528, 119736. The work features $(N, P, T)$ simulations and the densities reported could be higly impacted by the boron parameterization. Because one author is also a co-author of Ref.~[\onlinecite{Wang2018}], it is unclear whether the correct atomic masses were used in the simulations.

\item Lee, K.; Yang, Y.; Ding, L.; Ziebarth, B.; Davis, M. J.; Mauro, J. C. \href{https://doi.org/10.1111/jace.17681}{``Atomic-scale Mechanisms of Densification in Cold-compressed Borosilicate Glasses''}. \emph{J. Am. Ceramic Soc.}, \textbf{2021}, 104, 2506--2520. The same remarks as the previous work apply.

\end{itemize}

\section*{Data availability}

The simulation files used to reproduce the work (both with correct and incorrect atomic masses) are freely available at \url{https://github.com/fxcoudert/citable-data}

\begin{acknowledgments}
Mengyi Wang is acknowledged for sharing original simulation input files.
\end{acknowledgments}

\bibliography{article}

\end{document}